\documentclass[aps,superscriptaddress,amsmath,showpacs,twocolumn,pra]{revtex4-2}
\usepackage[T1]{fontenc}
\usepackage[latin9]{inputenc}
\usepackage{amssymb}
\usepackage{graphicx}
\usepackage{subfigure}
\usepackage{epsfig}
\usepackage{txfonts}
\usepackage{amsfonts}
\usepackage{dcolumn}
\usepackage{esint}
\usepackage{mathrsfs}
\usepackage{enumerate}
\usepackage{color}
\begin{document}

\title{Sensitively searching for microwave dark photons with atomic ensembles}
\author{Suirong He}
\affiliation{HgerD Collaboration and Information Quantum Technology Laboratory, School of Information Science and Technology, Southwest Jiaotong University, Chengdu 610031, China}
\author{De He}
\affiliation{HgerD Collaboration and Information Quantum Technology Laboratory, School of Information Science and Technology, Southwest Jiaotong University, Chengdu 610031, China}
\author{Yufen Li}
\affiliation{HgerD Collaboration and Information Quantum Technology Laboratory, School of Information Science and Technology, Southwest Jiaotong University, Chengdu 610031, China}
\author{Li Gao}
\affiliation{HgerD Collaboration and Information Quantum Technology Laboratory, School of Information Science and Technology, Southwest Jiaotong University, Chengdu 610031, China}
\author{Xianing Feng}
\affiliation{HgerD Collaboration and Information Quantum Technology Laboratory, School of Information Science and Technology, Southwest Jiaotong University, Chengdu 610031, China}
\author{Hao Zheng}
\affiliation{HgerD Collaboration and Information Quantum Technology Laboratory, School of Information Science and Technology, Southwest Jiaotong University, Chengdu 610031, China}
\author{L. F. Wei}
\email{Corresponding author: lfwei@swjtu.edu.cn}
\affiliation{HgerD Collaboration and Information Quantum Technology Laboratory, School of Information Science and Technology, Southwest Jiaotong University, Chengdu 610031, China}

\begin{abstract}
Dark photon is one of the promising candidates of light dark matter and could be detected by using its interaction with standard model particles via kinetic mixings. Here, we propose a feasible approach to detect the dark photons by nondestructively probing these mixing-induced quantum state transitions of atomic ensembles. Compared with the scheme by probing the mixing-induced quantum excitation of single-atom detector, the achievable detection sensitivity can be enhanced theoretically by a factor of $\sqrt{N}$ for the ensemble containing $N$ atoms. Specifically, we show that the dark photons, in both centimeter- and millimeter-wave bands, could be detected by using the artificial atomic ensemble detector, generated by surface-state electrons on liquid Helium.
It is estimated that, with the detectable transition probability of $10^{-4}$, the experimental surface-state electrons (with $N = 10^8$ trapped electrons) might provide a feasible approach to search for the dark photons in $18.61-26.88$ $\mu$eV and $496.28-827.13$ $\mu$eV ranges, within about two months. The confidence level can exceed 95\% for the achievable sensitivities being $10^{-14} \sim 10^{-13}$ and $10^{-12} \sim 10^{-11}$, respectively. In principle, the proposal could also be generalized to the other atomic ensemble detectors for the detection of dark photons in different frequency bands.
\end{abstract}
\maketitle

{\it Introduction.---}
The existence of dark matter~\cite{2018GB,2021AC} has been supported by various astronomic and cosmological observations, typically such as the cosmic microwave background radiation~\cite{2022EK}, gravitational lensing~\cite{2019SJ}, the bullet cluster~\cite{2015PWG}, and the large-scale evolution of universe~\cite{2006VS}, etc..
Particularly, the detection of dark photons (which serve as the hypothetical vector gauge bosons and possess similar characteristics of the ordinary photons) has been paid much attention~\cite{2021AC}. Although direct detection of dark photons remains elusive, constraints on their coupling to the visible matter have been given continuously by the relevant accelerator experiments and astrophysical observations~\cite{2015NV,2021XB}. To surpass these constraints, higher achievable detection sensitivities are required naturally.

Basically, dark photons can be detected by probing the potential interaction with the ordinary photons via the weak kinetic mixings~\cite{2021AC}. Indeed, a series of experiments typically including, the light shining through walls~\cite{2010RGP,2013SRP}, by using astronomical telescopes~\cite{2024HA}, and the high-quality resonant cavity techniques~\cite{2024RK,2021AVD}, etc..
Different from these telescopic techniques for the energy detection of the photons inverted from dark photons~\cite{2022FA,2023HA}, in recent years quantum sensing detections, i.e., probing the dark photon induced quantum transitions between the selected quantum states of certain single atomic detectors, have being proposed~\cite{2023SC,2019VVF,2019CAL}. The achievable detection sensitivity is expected to be enhanced by using the entanglement~\cite{2024AI} and squeezing techniques~\cite{2021KMB,2023AOS}, although their experimental implementations are still the great challenges~\cite{2004AC,2021GF}.
\begin{figure}[ht]
\centering
\includegraphics[width=6cm]{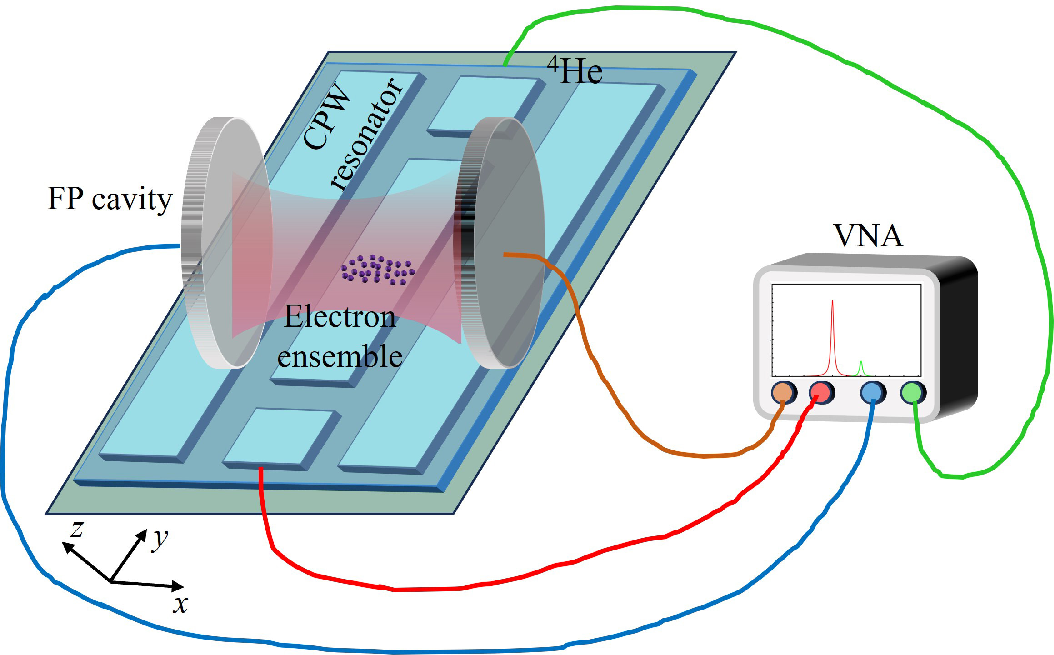}
\caption{An atomic ensemble detector configuration, generated by the surface-state electrons on liquid Helium, is designed for the sensitive detection of dark photons in the microwave band. The driven centimeter-wave resonator and the millimeter-wave Fabry-P\'{e}rot cavity are utilized to nondestructively probe the quantum state transitions, induced by the kinetic mixed dark photons in the relevant frequency bands, respectively.}
\end{figure}

Alternatively, in this letter, we propose a feasible approach, shown in Fig.~1, to implement the detection of dark photons in the microwave band, by using the atomic ensemble detectors generated by the surface-state electrons trapped on liquid Helium~\cite{2016GY,2018JC,2023EK}.
In fact, such an artificial atomic system possesses certain manifest advantages.
First, due to the relatively clean environment, the coherent times of the system could be significantly long (e.g., up to ms-order)~\cite{2023EK}. Second, the transition frequencies between two selected quantum states can be adjusted conveniently by controlling the biased voltages applied to the micro-electrodes at the bottom of the liquid Helium~\cite{2003MID}. As a consequence, the detection bandwidth of dark photons could be sufficiently enlarged. More interestingly, the transition between the selected quantum states can be nondestructively detected by driving dispersively coupled cavity~\cite{2023EK,2016GY,2018JC}. Therefore, such an experimental system~\cite{2016GY,2018JC} could be served as one of the promising candidates of atomic ensemble detector for the detection of dark photons.

{\it Model and Method.---}Theoretically, dark photons can be coupled weakly to standard model particles, through kinetic mixings~\cite{2021AC} described by the following effective Lagrangian~\cite{2019VVF,2023JG}:
$\tilde{\mathcal{L}}\supset-\tilde{F}_{\mu\nu}\tilde{F}^{\mu\nu}/4+ej^{\mu}(\tilde{A}_{\mu}-\chi A'_{\mu})-F'_{\mu\nu}F'^{\mu\nu}/4+M^{2}A'_{\mu}A'^{\mu}/2$,
with $\chi$ being the dimensionless mixed strength between dark photons and ordinary photons. Above, $j^{\mu}$ is the four-current of the ordinary field, $e$ is the electric charge of the electron, $M$ is the mass of dark photon, and $\tilde{F}_{\mu\nu}=F_{\mu\nu}+\chi F'_{\mu\nu}, F_{\mu\nu}=\partial_\mu A_\nu-\partial_\nu A_\mu$,  $F'_{\mu\nu}=\partial_\mu A'_\nu-\partial_\nu A'_\mu $ denote the relevant field strength tensors with $A_{\mu}$ and $A'_{\mu}$ being the four potentials of the ordinary and dark photons, respectively. Consequently, an additional electric field ${\bf E}(A'_\mu)=-\chi\sqrt{2\rho_{DM}}\vec{n}\cos(Mt)$, generated by the kinetic mixing between the dark- and ordinary photons, could be superposed effectively to the ordinary electric field ${\bf E}(A_\mu)$. Here, $\rho_{DM}$ is the density of dark photons, and $\vec{n}$ is the unit vector pointing in the direction of $A'_{\mu}$. With such a mixing, dark photons might lead to the effective interaction with the ordinary fermionic current $j^\mu$
\cite{2023SC}, i.e.,
$\mathcal{L}_{\mathrm{int}}=-e\bar{\Psi}_e\gamma^\mu\chi A'_{\mu}\Psi_e$, with $\Psi_e$ being
the fermionic wave function, thus resulting in a detectable effect.

Immediately, the interaction Hamiltonian, for a single two-state atom in the effective electric field ${\bf E}^{eff}(A_\mu, A'_\mu)={\bf E}(A_\mu)+{\bf E}(A'_\mu)$, can be expressed as $H^{eff}=-e{\bf d}_{eg}\cdot {\bf E}^{eff}(A_\mu, A'_\mu)$, where ${\bf d}_{eg}$ is the atomic dipole moment vector. As a consequence, by probing the additionally excited probability (under the electric dipole approximation)~\cite{2023SC,2024AI};
$P_{g\rightarrow e}(\tau)=\left|e{\bf d}_{eg}\chi\sqrt{2\rho_{DM}}\cos\Theta/(2\hbar)\right|^2 \sin c\left[(\omega_{eg}-M)\tau/2\right]^2\tau^2$,
of a single two-level atom being transited from the ground state $|g\rangle$ to the excited one $|e\rangle$~\cite{2023JG}, the dark photons could be detected effectively. Above, $\Theta\in(0,\pi)$ is the angle between the electric field vector, $e{\bf d}_{eg}$ is the electric dipole moment for a single atom, and $\omega_{eg}$ is the transition frequency of the atom. Indeed,
The feasibility of such an approach has been discussed by using the single superconducting transmon~\cite{2023SC} and trapped ion~\cite{2024AI}. The detection sensitivities, i.e., the detectable values of the mixing strength $\chi$ are expected to be $10^{-13}\sim 10^{-12}$ for $M\sim 4-40\mu$eV, and $10^{-13}\sim 10^{-12}$ for $M\sim 0.4-40$neV, respectively. Certainly, these sensitivities are still very low, due to the significantly weak interaction between single atoms and the mixed electric field.

To overcome such a difficulty, we now consider alternatively an atomic ensemble detector, illustrated schematically in Fig.~1, wherein the nonlinear vibrations of the surface-state electrons on liquid Helium along the $y$- and $z$-directions are dispersively coupled to the centimeter-wave resonator at the bottom of liquid Helium and the millimeter-wave cavity above the liquid Helium, respectively.
The basic idea by using the designed atomic ensemble detector to detect dark photons is as follows. Initially, the atomic ensemble, with $N$ atoms, is cooled at its ground state $|\psi_0\rangle=|g_1,g_2,\cdots,g_N\rangle$.
Next, the very weak mixing between dark photons and the ordinary photons (e.g., the  thermal noise photons) induces the desired excitation of the ensemble, under the interaction~\cite{2022SH,2013SC}
\begin{equation}
H^{eff}=-\sqrt{N}e\tilde{ d}_{eg}\cdot {\bf E}^{eff}(A_\mu, A'_\mu).
\end{equation}
Due to the simplest single-photon excitation within the duration $\tau$, with the probability $P_{\psi_0\rightarrow \psi_s}(\tau)=|\sqrt{N}e\tilde{ d}_{eg}{\bf E}^{eff}(A_\mu, A'_\mu)/(2\hbar)|^2\tau^2$, the atomic ensemble is generically evolved into the following single-excitation quantum superposed state: $|\psi_s\rangle=\sum_{j=1}^N|g_1,g_2,...e_j,...g_{N-1},g_N\rangle/\sqrt{N}$,
which is just the usual Werner state with single-excitation~\cite{2022SH,2013SC}. Above, $\tilde{ d}_{eg}=\sqrt{\hbar/2m\omega_{eg}}$ with $m$ being the mass of the atom. As a consequence,
the generic state of the atomic ensemble can be expressed as;
\begin{equation}
|\psi_g\rangle = \left(\sqrt{1-P_{\psi_0\rightarrow \psi_s}(\tau)}\right)|\psi_0\rangle + \sqrt{P_{\psi_0\rightarrow \psi_s}(\tau)}|\psi_s\rangle,
\end{equation}
due to the one-photon transition induced by both the ordinary photons and the mixed dark photons.
\begin{figure}[htbp]
\setlength{\abovecaptionskip}{0.cm}
\setlength{\belowcaptionskip}{-0.cm}
\centering
\includegraphics[width=6cm]{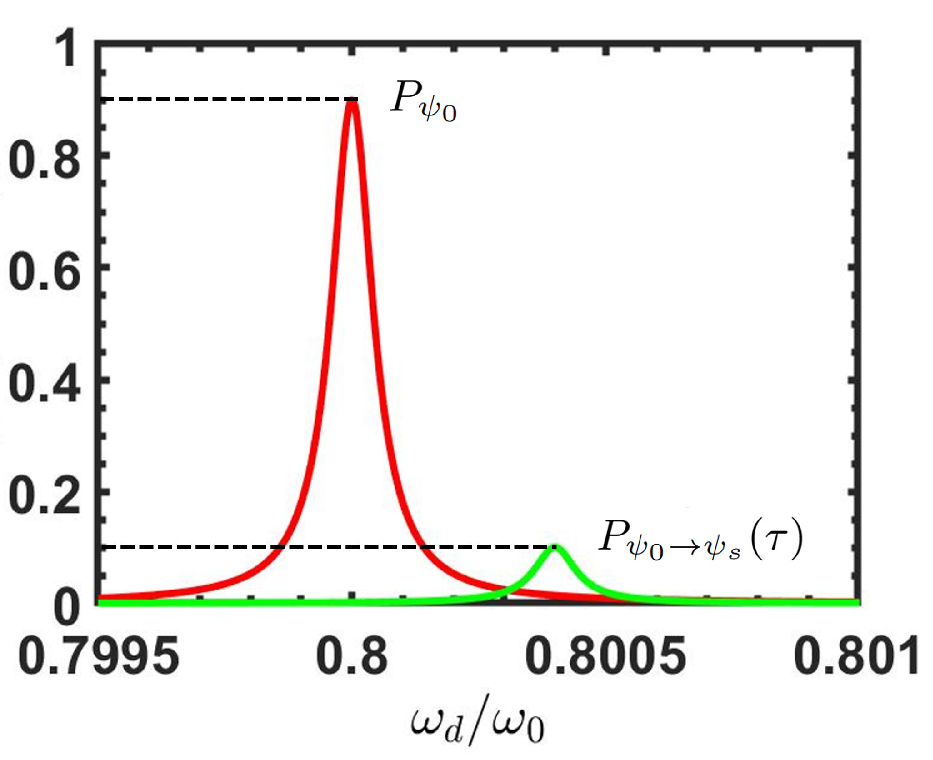}
\caption{The normalized steady-state transmitted spectrum of the driven cavity dispersively coupled to the atomic ensemble with $N$ atoms. The peaks of the red and green solid lines refer to the superposed probabilities of the atomic states being in $|\psi_0\rangle$ and $|\psi_s\rangle$, respectively. The relevant parameters are chosen as $\kappa=5\times 10^{-5}\omega_0, g=0.01\omega_0, \omega_{eg}=1.5\omega_0, N=1000$, $\varepsilon_d=10^2$V/cm, and $P_{\psi_0\rightarrow \psi_s}(\tau) = 0.1$, without loss of the generality.}
\end{figure}

Interestingly, the probability $P_{\psi_0\rightarrow \psi_s}(\tau)$ can be measured by using the driven cavity technique, i.e., probing the transmitting properties of the dispersively coupled cavity with the annihilation (creation) operator being $a$ ($a^\dagger$).
In fact, the transmitted spectrum of a driven cavity, dispersively coupled to an atomic ensemble, can be calculated as (see Appendix for details)~\cite{2016WG,2019SG}
\begin{equation}
\langle a^\dagger a\rangle (\omega_d)\propto\frac{\kappa^2+\Gamma^2-2\langle\psi_g|\sigma_{z}^{(j)}|\psi_g\rangle\Gamma\Delta_0+\Delta_0^2}{(\kappa^2+\Gamma^2-\Delta_0^2)^2+4\kappa^2\Delta_0^2},
\end{equation}
where $\kappa=\gamma/2$ is the cavity dissipation rate, $\Gamma= g^{2}/\Delta$ denotes the coupling strength between the atom and the cavity mode of frequency $\omega_0$, $\Delta = \omega_{eg}-\omega_{0}$ indicates the detuning between the atom and the cavity, $\varepsilon_d$ represents the strength of the driving field of frequency $\omega_d$, and $\Delta_0 = \omega_{0}-\omega'-\omega_{d}$ with $\omega'=(N-1)g^{2}/\Delta$. Above, we have assumed that all atoms in the ensemble share the same transition frequency $\omega_{eg}$ and coupling strength $g$ with the cavity.
In Fig.~2 we show specifically the spectrum of a driven cavity dispersively coupled to the atomic ensemble, wherein the red and green lines refer to the atomic ensemble being in the ground and one-excitation excited states, respectively. The value of $P_{\psi_0\rightarrow \psi_s}(\tau)$ can be measured by the relative height of the shifted peak. Certainly, besides the above frequency-domain measurements, the probability $P_{\psi_0\rightarrow\psi_s}(\tau)$ could also be measured by using the usual IQ-mixer in the time domain~\cite{2014EJ,2024YL}.

If the thermal noise, which shows the usual Planck blackbody spectrum with the fluctuated electric field~\cite{2022HZ}:
\begin{equation}
\begin{aligned}
\delta E_{them}=\sqrt{\overline{E^2}_{them}} &=\sqrt{\frac{2}{\epsilon_0}\int_0^\infty \frac{\hbar\omega_k^3}{\pi^2 c^3} \frac{1}{e^{\hbar\omega_k/k_B T}-1}d\omega_k},
\end{aligned}
\end{equation}
is served as the ordinary field of the present detector in temperature $T$, then the near-resonant excited probability of the atomic ensemble detector induced by the thermal radiation field could be calculated as:
\begin{equation}
\begin{aligned}
P^{them}_{\psi_0\rightarrow \psi_s}(\tau)&=\left|\frac{\sqrt{N}e\tilde{ d}_{eg}\delta E_{them}}{2\hbar}\right|^2\tau^2\\&=\frac{Ne^2 \tilde{d}_{eg}^2\tau^2}{2\epsilon_0\pi^2c^3\hbar}\int_{\omega_{eg}-\delta\omega}^{\omega_{eg}+\delta\omega}\frac{\omega_k^3}{e^{\hbar\omega_k/k_BT}-1}d\omega_k,
\end{aligned}
\end{equation}
within the frequency width $2\delta\omega$. Consequently, With the measured $P_{\psi_0\rightarrow \psi_s}(\tau)$ and Eqs.~(5), the sensitivity of the detection of dark photons can be calculated as
\begin{equation}
\begin{aligned}
\chi=\frac{\sqrt{2}\hbar}{\tau e\tilde{d}_{eg}\cos\Theta}\sqrt{\frac{P_{\psi_0\rightarrow \psi_s}(\tau)-P^{them}_{\psi_0\rightarrow \psi_s}(\tau)}{N\rho_{DM}}}\propto O(\frac{1}{\sqrt{N}}).
\end{aligned}
\end{equation}
Experimentally, the confidence level of detections is regarded as higher than $95\%$, if the criterion $S/\sqrt{B}>1.645$ is satisfied~\cite{2024AI}. Here, for  $N_{shot}$ observations, $B=N_{shot}P^{them}_{\psi_0\rightarrow \psi_s}(\tau)$ and  $S=N_{shot}[P_{\psi_0\rightarrow \psi_s}(\tau)-P^{them}_{\psi_0\rightarrow \psi_s}(\tau)$] are the numbers of the transitions detection caused by the thermal noise and the mixed dark photons, respectively. Physically, the excited duration $\tau$ of the atomic ensemble should be limited as
$\tau< min(\tau_{DP},\tau_{q})$, with $\tau_{q}$ and $\tau_{DP}=2\pi/M v_{DM}^2$ (where $v_{DM}\sim 10^{-3}$ is the relative speed of dark matter and can be estimated by the virial velocity of our galaxy~\cite{2024AI}) being the coherence time of the atomic ensemble and the dark photons, respectively.
Obviously, for a given detection probability $P_{\psi_0\rightarrow \psi_s}(\tau)$ within a fixed excited duration $\tau$ of the atomic ensemble in the fixed temperature $T$, the higher achievable detection sensitivity of dark photons corresponds to the narrower detection width $\delta\omega$ and the larger dipole momentum $\tilde{d}_{eg}$, although the signal-noise rate might be unchanged.

{\it Physical demonstrations with the electrons on liquid Helium.---} We now discuss how the detection sensitivity could be achieved by using the proposed detector configuration, shown in Fig.~1, to implement the detection of dark photons in both the centimeter- and millimeter bands.
\begin{figure}[htbp]
\setlength{\abovecaptionskip}{0.cm}
\setlength{\belowcaptionskip}{-0.cm}
\centering
\includegraphics[width=8.5cm]{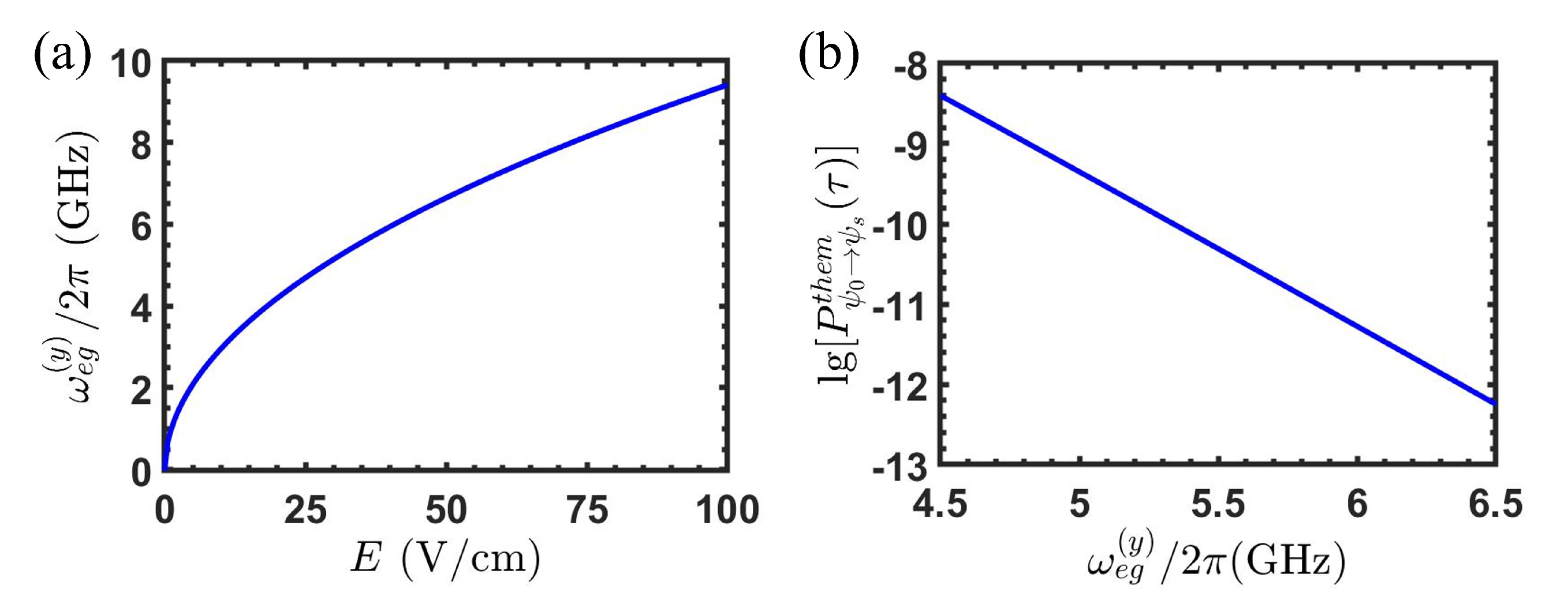}
\caption{(a) The adjustable eigenfrequency $\omega_{eg}^{(y)}$ of the electron vibrating nonliearly along the $y$-direction on liquid Helium, by controlling the biased electric field $E$. (b) The excited probabilitic spectrum $P^{them}_{\psi_0\rightarrow \psi_s}(\tau)$, induced by the thermal photons, within $\tau=10^{-4}$s in $T=10$mK.}
\end{figure}

Firstly, let us discuss the achievable detection sensitivity of detection of dark photons in the centimeter frequency band, by using the horizontal nonlinear vibrations along the $y$-direction of the electrons on liquid Helium, whose eigenfrequency $\omega_{eg}^{(y)}$ is in the centimeter band. Importantly, as shown in Fig.~3(a), such an atomic eigenfrequency is adjustable by controlling the external electric field, biased
via the micro-electrodes at the bottom of liquid Helium~\cite{2023EK,2024YL}. For example, we have $\omega_{eg}^{(y)}/2\pi \approx 5.143$ GHz, if the biased electric field $E$ is set as $30$ V/cm and the distance between the liquid Helium surface and the micro-electrodes is set as $l = 0.5\mu\text{m}$, respectively. In Fig.~3(b) we show the calculated the transition probability $P^{them}_{\psi_0\rightarrow \psi_s}(\tau)$ of the atomic ensemble induced by thermal photons in the frequency range of $4.5$ to $6.5$ GHz with the frequency broadening $\delta \omega = 1$ kHz for the electron density on liquid Helium ~\cite{2016GY} being approximately $10^{8} \text{cm}^{-2}$, within the coherence time window of the atoms being $\tau = 10^{-4}$s (i.e., the detection duration) in $T=10$ mK. This implies that the confidence level of the desired dark photons detection can exceed $95\%$, if $P_{\psi_0\rightarrow \psi_s}(\tau)\geq 10^{-4}$ is obtained
by probing the transmission of the driven transmission line resonator at the bottom of liquid Helium ~\cite{2016GY}, which is dispersively coupled to the atomic ensemble. Accordingly, the detection sensitivity of the dark photons in this band can be expressed as~\cite{2023JG}
\begin{equation}
\begin{aligned}
\chi_{cm}=& \frac{1.075\times10^{-12}}{\cos\Theta}\times\left(\frac{P_{\psi_0\rightarrow \psi_s}(\tau)-P^{them}_{\psi_0\rightarrow \psi_s}(\tau)}{10^{-4}}\right)^{\frac{1}{2}}\\
&\times\left(\frac{N}{10^5}\right)^{-\frac{1}{2}}\times\left(\frac \tau{10^{-4}s}\right)^{-1}\times\left(\frac m{0.51\text{MeV}}\right)^{\frac12} \\
&\times\left(\frac{\omega^{(y)}_{eg}}{2\pi\times1\text{GHz}}\right)^{\frac12}\times\left(\frac{\rho_{DM}}{0.45\text{GeVcm}^{-3}}\right)^{-\frac12},
\end{aligned}
\end{equation}
which can reach $\sim 10^{-14}-10^{-13}$ for $10000$ observations within $10$s by using the resonator with the quality factor being $Q=10^6$, corresponding to the bandwidth is $5.5$KHz.
\begin{figure}[htbp]
\setlength{\abovecaptionskip}{0.cm}
\setlength{\belowcaptionskip}{-0.cm}
	\centering
\includegraphics[width=6cm]{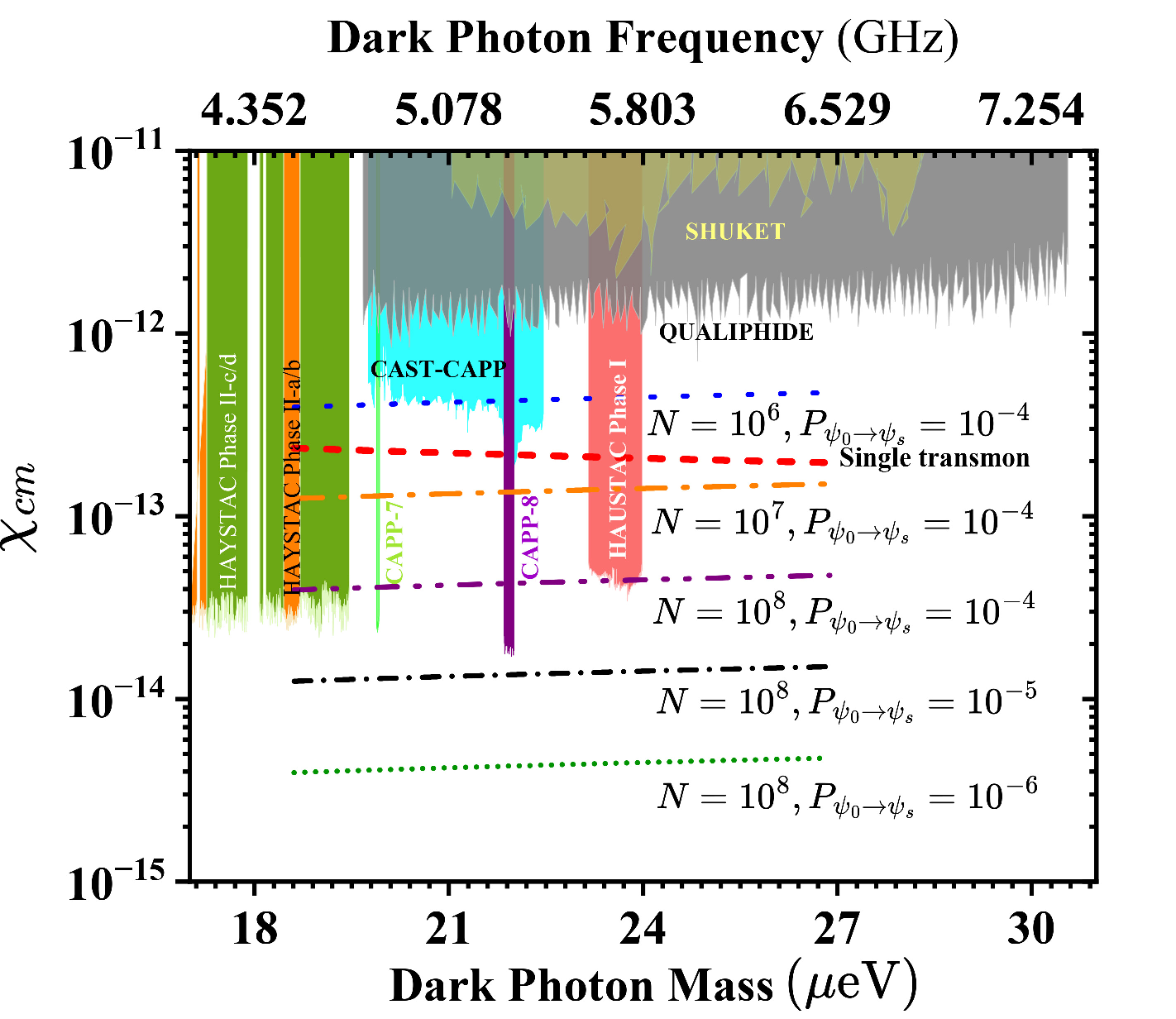}
\caption{The achievable detection sensitivities of dark photons in centimeter wave band by using the atomic ensemble detector, generated by the $y$-direction nonlinear vibration of electrons on liquid Helium. Here, we set   $\tau=10^{-4}$s and $N=10^6, 10^7$, and $10^8$ for $P_{\psi_0\rightarrow \psi_s}=10^{-4}$, $10^{-5}$ and $10^{-6}$, respectively. The other parameters are set as; $\rho_{DM} = 0.45$ GeV/cm$^{3}$ and $\cos^2\Theta = 1/3$, $T=10$mK.}
\end{figure}
Therefore, one can estimate that the total duration, for the scanning detection of dark photons within the mass range being $18.61\mu\text{eV}$-$26.88\mu\text{eV}$, would be about 42 days. As shown by the black (deep green) line in Fig.~4 that, if we set $P_{\psi_0\rightarrow \psi_s}(\tau)= 10^{-5}$ ($10^{-6}$), the total detection duration should be set as $T_d=100$s ($1000$s) for single frequency points, and then the achievable detection sensitivity can be improved as $\chi_{cm}\sim 10^{-14} (\sim 10^{-15}$). Compared with the other detection schemes~\cite{2021AC}, the present atomic ensemble detector (with $N=10^8$ artificial atoms) possesses certain sensitive advantages typically over the CAPP-8~\cite{2021AC}, for $P_{\psi_0\rightarrow \psi_s}(\tau)\leq 10^{-5}$ and the proposal by using a superconducting transmon~\cite{2023SC}, for $P_{\psi_0\rightarrow \psi_s}(\tau)\leq 10^{-4}$, respectively.

\begin{figure}[htbp]
\setlength{\abovecaptionskip}{0.cm}
\setlength{\belowcaptionskip}{-0.cm}
\centering
\includegraphics[width=8.5cm]{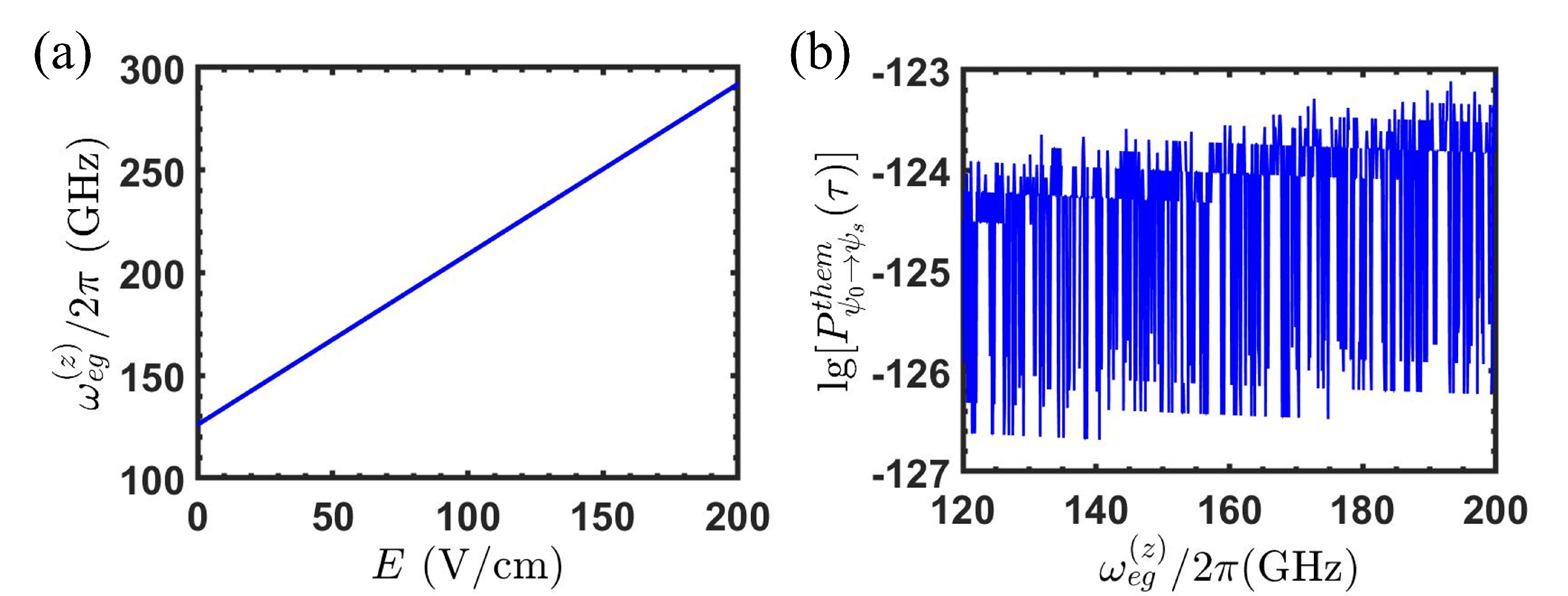}
\caption{(a) The adjustable eigenfrequency $\omega_{eg}^{(z)}$ of the trapped electron vibrating along the $z$-direction by controlling the biased electric field $E$. (b) The thermal one-photon excited probability $P^{them}_{\psi_0\rightarrow \psi_s}$ of the atomic ensemble in the millimeter wave band in $T=10$mK.}
\end{figure}

Importantly, the same experimental system can also be utilized to simultaneously search for the dark photons in the millimeter-wave band, if the nonlinear vibration of the trapped electrons along the $z$-direction are also served as another atomic ensemble detector, whose feasible eigenfrequency being adjustably in the frequency band; $\omega_{eg}^{(z)}\sim 120-280$GHz, shown in Fig.~5(a). The nondestructive readout of the present atomic state shown similarly in Eq.~(2) is achieved by driving the dispersively coupled Fabry-P\'{e}rot cavity above the liquid Helium, alternatively.
\begin{figure}[htbp]
\setlength{\abovecaptionskip}{0.cm}
\setlength{\belowcaptionskip}{-0.cm}
\centering
\includegraphics[width=6cm]{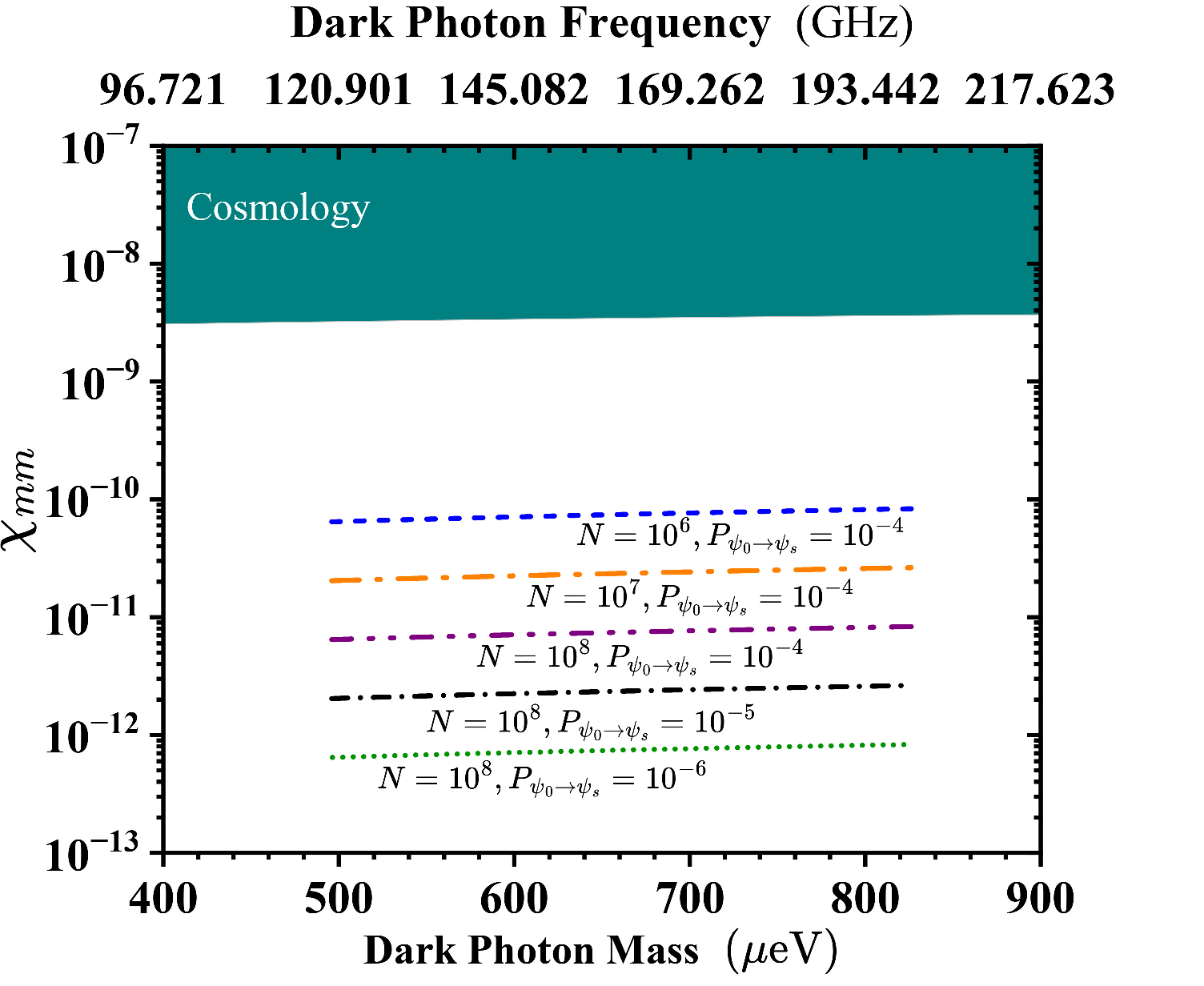}
\caption{The achievable detection sensitivity of dark photons in millimeter wave band by using the electron ensemble on liquid Helium. The shaded area indicates limits given by the cosmological observations~\cite{2021AC}. The blue, orange, and purple lines refer to those obtained by using the relevant atomic ensemble with the atomic numbers $N=10^6, 10^7$, and $10^8$, respectively for the transition probabilities of $P_{\psi_0\rightarrow \psi_s}(\tau)=10^{-4}$ and $\tau=10^{-5}$s. The black and deep green lines corresponds to $P_{\psi_0\rightarrow \psi_s}(\tau)=10^{-5}$ and $10^{-6}$, respectively. The other parameters are the same as in Fig.~4.}
\end{figure}
As shown in Fig.~5(b), the transition probability from $|\psi_0\rangle$ to $|\psi_s\rangle$, induced by the thermal photons in this frequency range is less than $\sim 10^{-125}$, for $N =10^8$, $T = 10$ mK, $\delta \omega = 1$ kHz, and $\tau = 10^{-5}$s. This implies that, the confidence level for detecting dark photons in this frequency band can exceed $95\%$, for the desired $P_{\psi_0 \rightarrow \psi_s}(\tau) \geq 10^{-4}$. Accordingly, with the replacement of the $\omega_{eq}^{(y)}$ in Eq.~(7) by $\omega_{eq}^{(z)}$, one can see that the achievable detection sensitivity can reach $\chi_{mm} \sim 10^{-12}$-$10^{-11}$ within $10$ s for $10000$ observations, by using the Fabry-P\'{e}rot cavity with the quality factor being $Q = 10^5$ (i.e., the present detection bandwidth becomes $1.6$ MHz). Therefore, the total detection duration required for scanning and detecting the dark photons within the mass range; from $496.28 \mu$eV to $827.13 \mu$eV, is estimated as about $58$ days. Notably, as shown by the black (deep green) line in Fig.~6, if the scanning duration for each detection frequency point is increased as $T_d= 100$s ($1000$s), the detection sensitivity can be further improved as $10^{-12}$ ($10^{-13}$), for $P_{\psi_0 \rightarrow \psi_s}(\tau) \geq 10^{-5}$ ($10^{-6}$) and $N=10^8$. Compared with the above detection in the centimeter wave band, the achievable detection sensitivity is decreased by about two orders of magnitude, due to the higher transition frequency and lower coherent time.
Anyway, compared with the cosmological observations~\cite{2021AC}, the achievable detection sensitivity implemented by using the present atomic ensemble detector could still be enhanced by two orders of magnitude, at least.

{\it Conclusions and discussions.---}
In summary, we propose a feasible approach to search for the dark photons in microwave band, by using the experimental atomic ensemble detector. Due to the collective coherent effect of the atomic ensemble, the detection sensitivity could be enhanced by a factor of $\sqrt{N}$, compared to that by using the corresponding single atom detector. The key technique to achieve such a detection is the nondestructive readout of the one-excitation state of the atomic ensemble detector, by driving the dispersively coupled cavity. Compared with the other detection schemes, we argue that the achievable detection sensitivity of the present detector could be enhanced by an order of magnitude, at least, and could be further improved by increasing the readout accuracy of the one-excitation state of the proposed atomic ensemble detector.

Although most of the experimentally atomic ensembles, typically such as NV centers in diamond~\cite{2020JFB}, Rydeberg atoms~\cite{2019SG}, and the Bose-Einstein condensed gases~\cite{2002JRA}, etc., can be also served as these ensemble detectors, in the present work we perfectly used the artificial atomic ensemble generated by electrons on liquid Helium. Our numerical estimations indicate that utilizing such an atomic ensemble detector containing $N\sim 10^8$ artificial atoms, one can implement the searches of the dark photons in the frequency range of $4.5\sim 6.5$ GHz (the mass range is $18.61\,\mu\text{eV} \sim 26.88\,\mu\text{eV}$) and simultaneously $120\sim 200$ GHz ($496.28\,\mu\text{eV} \sim 827.13\,\mu\text{eV}$), within 42 days and 58 days, respectively. At the confidence level over $95\%$, the achievable detection sensitivities can be arrived as $10^{-14} \sim 10^{-13}$ and $10^{-12} \sim 10^{-11}$, respectively.
Given such an artificial atomic ensemble had been experimentally demonstrated~\cite{2016GY,2018JC}, it is expected that the proposed atomic ensemble detector could be applied to experimentally search for the dark photons in the microwave band, in the near future.

\section*{Acknowledgments}
This work was partially supported in part by the National Key Research and Development Program of China under Grant No.~2021YFA0718803, the National Natural Science Foundation of China under Grant No.~11974290, and the Fundamental Research Funds for the Central Universities under Grant No.~2682024CX048.

\appendix
\section*{Appendix}
In this Appendix, we provide the derivation for the steady-state photon transmission spectrum of the driven cavity dispersive coupled to the atomic ensemble in state (2). The Hamiltonian of the present system reads $H=H_{N}+H_d$,
where
\renewcommand\theequation{A1}
\begin{equation}
H_{N}=\hbar\omega_{0}a^{\dagger}a+\sum_{j=1}^{N}\left[\frac{\hbar\omega_{eg}}{2}\sigma_{z}^{(j)}+\hbar g(a\sigma_{+}^{(j)}+\sigma_{-}^{(j)}a^{\dagger})\right]
\end{equation}
represents the interaction of the identical atoms with the cavity, $g$ and $\omega_{0}$ are respectively the cavity-atom coupling strength and frequency of the driven cavity, described by~\cite{2016WG,2013SC}.
\renewcommand\theequation{A2}
\begin{equation}
H_d=\hbar\varepsilon_d\big(a^\dagger e^{-i\omega_dt}+ae^{i\omega_dt}\big).
\end{equation}
Here, $\varepsilon_d$ and $\omega_d$ are the amplitude and frequency of the driving field. In the rotating frame defined by $e^{-it\omega_d\sigma_z^{(j)}}$, the Hamiltonian of system can be rewritten as~\cite{2019SG}:
\renewcommand\theequation{A3}
\begin{equation}
\frac{\tilde{H}}{\hbar}=\sum_{j=1}^N\frac{\tilde{\omega}_{eg}}{2}\sigma_{z}^{(j)}+(\Delta_0+\Gamma\sigma_{z}^{(j)})a^\dagger a+\varepsilon_d(a^\dagger+a),
\end{equation}
where $\tilde{\omega}_{eg}= \omega_{eg}+\Gamma$, $\Gamma = g^{2}/\Delta$, and $\Delta = \omega_{eg}-\omega_{0}$ is the detuning between the transition frequency of the atom and the resonant frequency of the cavity. Additionally, $\Delta_0 = \omega_{0} + \sum_{j=1}^{N-1} g^{2} \langle\sigma_{z}^{(j)}\rangle/\Delta-\omega_{d}$.

Generally, the dynamics of the system can be described by the following master equation~\cite{2016WG}: \renewcommand\theequation{A4}
\begin{equation}
\begin{aligned}
    \dot{\rho}=&-i[\tilde{H},\rho]\\
    &=\frac{1}{2}\gamma(\bar{n}+1)\big(2a\rho a^{\dagger}-a^{\dagger}a\rho-\rho a^{\dagger}a\big)\\&+\frac{1}{2}\gamma \bar{n}\big(2a^{\dagger}\rho a-aa^{\dagger}\rho-\rho aa^{\dagger}\big)
\end{aligned}
\end{equation}
where $\gamma$ is the cavity dissipation rate and $\bar{n}=1/(e^{\hbar\omega_0/k_B T}-1)$ is the thermal average photon number of the cavity mode, with $k_B$ being the Boltzmann constant and $T$ the temperature.
Substituting Eq.~(A3) into (A4), we have \renewcommand\theequation{A5}
\begin{equation}
\frac{d\langle a^\dagger a\rangle}{dt}=-i\varepsilon_d(\langle a^\dagger\rangle-\langle a\rangle)-\gamma\langle a^\dagger a\rangle+\gamma \bar{n},
\end{equation}
\renewcommand\theequation{A6}
\begin{eqnarray}
\left\{ \begin{aligned}
&\frac{d\langle a\rangle}{dt}=-i\Delta_0\langle a\rangle-i\Gamma\langle a\sigma_{z}^{(j)}\rangle-i\varepsilon_d-\frac{\gamma}{2}\langle a\rangle,\\
&\frac{d\langle a\sigma_{z}^{(j)}\rangle}{dt}=-i\Delta_0\langle a\sigma_{z}^{(j)}\rangle-i\Gamma\langle a\rangle-\frac{\gamma}{2}\langle a\sigma_{z}^{(j)}\rangle-i\varepsilon_d\langle\sigma_{z}^{(j)}\rangle,\\
&\frac{d\langle a^\dagger\rangle}{dt}=i\Delta_0\langle a^\dagger\rangle+i\Gamma\langle a^\dagger\sigma_{z}^{(j)}\rangle+i\varepsilon_d-\frac{\gamma}{2}\langle a^\dagger\rangle,\\
&\frac{d\langle a^\dagger\sigma_{z}^{(j)}\rangle}{dt}=i\Delta_0\langle a^\dagger\sigma_{z}^{(j)}\rangle+i\Gamma\langle a^\dagger\rangle+i\varepsilon_d\langle\sigma_{z}^{(j)}\rangle-\frac{\gamma}{2}\langle a^\dagger\sigma_{z}^{(j)}\rangle,
\end{aligned} \right.
\end{eqnarray}
and
$d\langle\sigma_{z}^{(j)}\rangle/dt=0$.
By solving Eqs.(A5-A6), the transmitted spectrum of the driven photons through the dispersively coupled cavity can be calculated as
\renewcommand\theequation{A7}
\begin{equation}
\langle a^\dagger a\rangle (\omega_d)=\varepsilon_d^2\frac{\kappa^2+\Gamma^2-2\langle\psi_g|\sigma_{z}^{(j)}|\psi_g\rangle\Gamma\Delta_0+\Delta_0^2}{(\kappa^2+\Gamma^2-\Delta_0^2)^2+4\kappa^2\Delta_0^2}+\bar{n}.
\end{equation}
Neglecting the last constant term, we get Eq.~(3).

\end{document}